\documentclass[aps,preprint]{revtex4}
\usepackage{graphicx}

\begin{document}

\title{Expansion of image space in enhanced-NA Fresnel holographic display}

\author{Byung Gyu Chae}

\address{Holographic Contents Research Laboratory, Electronics and Telecommunications Research Institute, 218 Gajeong-ro, Yuseong-gu, 
Daejeon 34129, Republic of Korea}


\begin{abstract}

The enhanced-NA Fresnel hologram reconstructs a holographic image at a viewing angle larger than the diffraction angle of a hologram pixel.
The image space is limited by the bandwidth of a digital hologram.
In this study, we investigate the property of image formation in the extended image space beyond a diffraction zone.
A numerical simulation using the phase Fresnel hologram is carried out to observe an extension of image space and the effect of this on the changes in the angular field of view.
The phase Fresnel hologram, synthesized by restricting the angular view range to a diffraction angle, can reconstruct a uniform image without high-order noises within the primary viewing zone,
which is well confirmed by optical experiments. 
On the other hand, the overlapping of high-order images is inevitable when the viewing angle depends on the hologram numerical aperture.
The high-order images are distributed in the direction cosine space, which could be effectively removed through an angular low-pass filter.
We discuss the development of method for expanding the image space while maintaining the viewing angle of a holographic image.

\end{abstract}

\maketitle{}

\section{Introduction}
A holographic display offers a three-dimensional image in such a way that the wavefront of the propagating wave from the digital hologram is generated in free space \cite{1}.
Various attempts to implement holographic displays have been conducted because it has the ultimate merit in that an observer can view a complete scene without visual fatigue \cite{2,3,4}.
Despite technological advancements,
there still remains a fundamental limitation for realizing a commercial holographic display owing to the finite space-bandwidth product of a pixelated spatial light modulator (SLM).

In particular, the issue of solving a narrow viewing angle is a crucial task.
It is known that an SLM with a pixel pitch of the wavelength scale is necessary for seeing a holographic image at a wide angular view.
A commercial modulator has a pixel pitch of several micrometers, inducing a viewing angle of a few degrees.
Experiments to overcome this problem have been conducted by expanding the diffraction zone using spatial or temporal multiplexing of modulators \cite{3,5,6,7,8}.
In this circumstance, massive data processing is required, and it would be inefficient in view of energy consumption in informatics.

These conventional approaches for expanding the viewing zone are based on the assumption that the diffraction angle of a pixel pitch directly relates to the angular view of the holographic image.
Previously, we found that the viewing angle $\it{\Omega_V}$ of the reconstructed image is fundamentally determined by numerical aperture (NA) of the digital hologram rather than a diffraction angle,
as demonstrated by the following equation \cite{9,10}:
\begin{equation}
\it{\Omega_V} = \rm{}2 \sin^{-1} \left(\rm{NA} \right).
\end{equation}
The hologram numerical aperture is represented as a geometrical structure of aperture size, $l = N\it{\Delta} x$, and a distance $z$, $\rm{NA} = \it{N\Delta} x /\rm{2}\it{z}$,
where $N$ and $\it{\Delta} x$ denote the number of pixels and pixel size of the digital hologram, respectively.
The viewing angle does not independently depend on the diffraction ability of the pixel pitch or the illuminating wave.
This concept can provide a major technological breakthrough in expanding the viewing angle in holographic display system.
In principle, the viewing angle varies with the reconstructed distance, even when using an SLM with a constant pixel pitch.
The viewing angle of the image would be larger than a diffraction angle when the image is formed below the critical distance.
We called this system the enhanced-NA Fresnel holographic display,
where there is still a decrease in image size because of the Nyquist sampling criterion in hologram synthesis \cite{9,11,12,13}.
However, this behavior is different from the trade-off relation characterized by the space-bandwidth product of the digital hologram itself.
For the realization of a wide viewing-angle holographic display, a holographic image without sacrificing the image size should be reconstructed.
When the image reconstruction is numerically or optically processed on the region outside of the sampling condition, the interference of high-order images is inevitable.
Therefore, a detailed study on the expansion of image space and the effect of this on the change in the viewing angle is required.

A phase modulation has some advantages, such as showing no twin image and high diffraction efficiency \cite{14,15,16}.
A phase retrieval algorithm such as the iterative Fresnel transform is used to calculate the phase hologram in the Fresnel diffraction regime.
The enhanced-NA digital hologram creates replica fringe patterns, 
which fundamentally obstructs the expansion of image space, beyond the bounds of a diffraction scope of a hologram pixel \cite{9}.

In this study, we explore the image reconstruction in the extended image space from the phase Fresnel hologram with a high numerical aperture.
Firstly, the phase hologram was successfully prepared using the iterative algorithm based on the Fresnel propagation.
We analyzed an angular view of a holographic image dependent on the hologram numerical aperture, and an expansion of image space.
Secondly, optical experiments were conducted to obtain the reconstructed image without an interference of high-order noises.
Subsequently, we studied the method for reconstructing a holographic image with an extended image volume at a larger viewing angle.

\section{Holographic image reconstructed from enhanced-NA phase Fresnel hologram}
\subsection{Analysis on angular field of view of holographic image dependent on hologram numerical aperture}
The Fresnel field, which propagates from the digital hologram to image plane, is described by the Fresnel diffraction formula.
The specifications of both planes are determined by the Nyquist-Shannon sampling theorem \cite{10,11,12}.
The relationship between pixel pitches $(\it{\Delta} x, \it{\Delta} y)$ and $(\it{\Delta} x', \it{\Delta} y')$ in the hologram $(x,y)$ and image $(x',y')$ coordinates is represented as follows:
\begin{equation}
\it{\Delta} x = \frac{\lambda z} {N\it{\Delta} x'},  \hspace{3mm} \it{\Delta} y = \frac{\lambda z} {N\it{\Delta} y'}.
\end{equation}
For convenience, one-dimensional description is considered hereafter.
The sampling condition of both planes is characterized by the critical distance having a value, $z_c = N \it{\Delta} x^{\rm{2}} /\lambda$
when the pixel pitches of both planes are the same.
The image size, $l' = N\it{\Delta} x'$ is smaller than the hologram size, $l = N\it{\Delta} x$ when the image is placed at a distance less than $z_c$.
We know that aliased errors appear if they are digitally or optically processed in the regions that deviate from this specification.

A practicable phase hologram $\phi$ for the real-valued object is extracted from the modified Gerchberg-Saxton (GS) iterative algorithm \cite{14,17,18}.
Figure 1 is a flow chart of the modified GS algorithm to obtain the phase-only hologram.
The Fresnel transform, consisting of a single Fourier transform, is used in iterative procedures.
Only the phase term is captured after the Fresnel transform, where the constraint in the spectral domain is a uniform modulus.
The inverse Fresnel transform of the exponential term $e^{i\phi}$ is carried out to modulate the phase hologram.
To repeat this process, the amplitude of the reconstructed image is replaced by the object value.
The algorithm rapidly converges to an optimized solution.

The phase holograms, composed of 1920×1920 pixels with an 8-µm pixel pitch, are synthesized using a letter image, as depicted in Figure 2(a).
Here, the critical distance considering a wavelength of $\lambda$=473 nm is calculated to be 259.8 mm.
To measure the viewing angle of a holographic image, the propagating wave far away from the reconstructed image was simulated.
 Figure 2(b) shows the propagating waves in the simulation using the digital hologram made at a critical distance.
The intensity profiles of diffractive wave indicated by a lateral line in inset picture are displayed with distance.
An observer can view the image within the spreading angle of diffractive wave.
From this, the viewing angle of the reconstructed image is estimated to be 4.2$^\circ$.
Figure 2(c) is the simulation result for a digital hologram made at a half of the critical distance.
The propagating wave from the image largely spreads, where the viewing angle appears to be about 7.7$^\circ$.

As illustrated in Fig. 2(a), a holographic image is reconstructed by illuminating the digital hologram with a plane wave.
Considering a point object,
it appears that the point image is focused on a particular place through the digital hologram acting as a lens. 
From Eq. (2), we can deduce that the lateral size and viewing angle $\it{\Omega_V}$ of the reconstructed image have a trade-off relation,
which is unlike the interpretation found by controlling the size of hologram itself.
The viewing angle is larger than the diffraction angle by a hologram pixel, $\theta = \rm{2} \sin^{-1} (\frac{\lambda} {2\it{\Delta} x})$ 
when the image is formed at a distance less than the critical distance.
The high-definition image diffracts to a larger spreading angle.
We know that the angle value $\it{\Omega_V}$ is pertinent to the hologram numerical aperture, $\rm{NA}= \it{n} \sin (\it{\Omega} /\rm{2})$,
where the refractive index $n$ is equal to one in free space.
Hence, the viewing angle of the reconstructed image obeys the relation in Eq. (1).
The above-measured viewing-angles are consistent with the calculated values.
We have also confirmed the viewing-angle variation dependent on the hologram numerical aperture in optical experiments \cite{9}.

Meanwhile, a method including a double Fourier transform such as convolutional algorithm or angular spectrum method cannot enhance the angular view,
because spatial filtering at the Fourier plane within the formula is embedded, resizing the hologram pattern according to the distance \cite{10}.
Here, one preserves the constant NA irrespective of the distance.
For clarity, we defined a hologram that maintains a high numerical aperture as the enhanced-NA Fresnel hologram \cite{13}.

In previous work, we found that an aliased error arises when synthesizing the enhanced-NA Fresnel hologram
because of the undersampling of the point spread function $h(x,y)$ in the front of Fourier transform $FT$ in the expanded form of the Fresnel diffraction formula, which is as follows \cite{13,19}:
\begin{equation}
h(x,y)FT\left[o(x',y')h(x',y')\right].
\end{equation}
Since the sampling criterion with respect to the object plane is well satisfied, the spreading angle of diffractive wave from the object $o(x',y')$ covers the whole area of digital hologram,
which induces the suppression of aliased replica fringes when the object has a finite size.
We explained that this phenomenon results from the near-field diffraction property within the diffraction zone by object pixel resolution \cite{9}.

This behavior brings about an energy concentration in a low-frequency region in the digital hologram, which weakens the diffraction ability by an incident plane wave.
Figure 3 illustrates the diffractive wave from the reconstructed image by a directly calculated digital hologram.
The simulated light intensity from the finite-sized image does not spread out with distance,
where it is difficult to observe a holographic image with a sufficient angular view.
On the other hand, in a logarithmic scale, the spreading angle of diffractive wave is clearly distinguished \cite{10}.
A holographic image floating in free space is induced from part of digital hologram,
while most of concentrated hologram fringes directly propagates without their diffraction.  
An effective method for mitigating the energy concentration is to add a random phase to the object in the hologram synthesis.
In this case, as depicted in Fig. 3(c), the intensity profile of diffractive wave progresses at an adequate spreading angle.

\subsection{Expansion of image space and its reconstruction property depending on viewing angle}
For the realization of a viable holographic display based on the enhanced-NA Fresnel hologram, a sufficient amount of image space should be secured.
Figure 4 shows the simulation results for the generation of high-order images in the image space.
The digital hologram with an 8-µm pixel pitch was conveniently calculated by using a rectangular object consisting of 256×256 pixels, and then, the pixels are upsampled tenfold.
The pixelated structure plays a role in planar gratings, and the modulated replica images were generated at an interval of $\lambda z/p$ in the Fresnel regime, where $p$ indicates a pixel interval \cite{10}.
In this specification, it is expected that the Fraunhofer pattern starts to appear at a submillimeter distance.
The digital hologram made at a distance $2z_c$, 69.3 mm reveals the enveloped images by a sinc function.
However, there is no intensity modulation in the reconstructed image from the enhanced-NA Fresnel hologram,
although the image is placed at a millimeter distance.
Here, the bright areas of all the images are comparable to hologram size on the same scale.
The image intensity has been known to be uniform irrespective of its lateral location \cite{20,21}.
This makes it possible to reconstruct a uniform image in the extended image space larger than the diffraction scope by a hologram pixel.

Figure 5 displays the schematics of image reconstruction in the extended image space.
We consider the hologram made using the letter object placed at a distance of half of $z_c$, 17.3 mm.
The object has 512×512 pixels with a 4-µm pixel pitch, which is 2 times larger than the diffraction scope by a hologram pixel pitch.
The angular separation of diffracted orders from the planar gratings is well described in the direction cosine space \cite{22,23}.
Each diffraction vector carries the high-order image in the new direction.
As depicted in Fig. 5(a), the first-order images are created in the corresponding areas outside of the diffraction scope.
The viewing direction of high-order image is tilted at a diffraction angle $\theta_m$, which is shown in the following equation:
\begin{equation}
\sin \theta_m = \frac{m \lambda} {p}.
\end{equation}
Figure 5(b) is the numerically reconstructed image showing up to the second-order term.
The Rayleigh-Sommerfeld formula was used to preserve non-paraxial diffraction.
The high-order images are unfocused because the image plane vertical to optical axis is slightly apart from the direction cosine surface.
The image blur is larger at higher order terms.

We characterize the reconstruction behavior in two separate cases.
The first case is that the viewing angle depends on the hologram numerical aperture, whose value is double a diffraction angle in this configuration.
Here, the overlapped image with the high-order terms is observable.
This type of high-order images would be removed through an angular low-pass filter, which will be discussed later.
Secondly, the viewing angle could be restricted to a diffraction angle due to the nature of algorithm.
In this case, the complete image without the interference of high-order noises can be seen in the viewing direction of the zeroth order.
If the viewing position is shifted to the first-order direction, the complete replica image will still reappear.
This phenomenon is well observed in optical experiments.

\section{Optical reconstruction from enhanced-NA Fresnel hologram without sacrificing image size}

We carried out optical experiments by using a phase-only SLM (Holoeye PLUTO) with 1920×1080 pixels of an 8-µm pixel pitch and a blue laser with a wavelength of 473 nm.
The critical distance for a pixel specification in a vertical direction was set as a reference to secure the imaging area with no aliased errors, which was calculated to be 146.1 mm.
The phase holograms were synthesized using a cameramen image located at distances of a half, a quarter, and one eighth of $z_c$, in Fig. 6(a).
It is hard to directly use the fast Fourier transform algorithm in the hologram synthesis using the object space outside of the diffraction scope
because the pixel specifications of both planes deviate from the sampling condition in Eq. (2).
The digital hologram was prepared in two separate steps.
First, the high-resolution holograms used to suit the extended object space in accordance with the sampling condition were calculated using the modified GS algorithm;
thereafter, they were downsampled to the final resolution of the hologram.

To prepare a hologram made at half $z_c$, 73.1 mm,
the initial phase hologram with 3840×2160 pixels of a 4-µm pixel pitch was calculated using the object with the same pixel specifications.
The final hologram in Fig. 6(b) is the downsampled data by twice the pixel interval.
We conveniently captured the image within the SLM size, by considering a direct-beam filtering by a polarizer.
Figure 6(c) shows the reconstructed image with no replica noises.
We observed that the other holograms reconstructed the original images well, without suffering from the high-order noises.
All the images have a rectangular form with an 8.64 mm×8.64 mm lateral size, according to the sampling condition.
Although it is not displayed here, when the viewing direction is moved to the high-order region, the corresponding replica image reappears after the primary image is gone,
which is consistent with the numerical analysis.
The high-order images that overlapped in the image space are not viewable because of an inclined viewing direction.
The image space is expanded, but the viewing angle is confined to a diffraction angle.

Figure 6(d) is the unfocused image about the image of Fig. 6(c).
Even though an image blurring occurs, an accommodation effect is very weak.
This indicates that the diffractive wave from the image spot propagates at a narrow spreading angle, where a perspective view of the image is hardly visible to the naked eye.
The energy compaction of hologram patterns invokes this weakness, as explained in Section 2.

In second experiment, a sparse letter object was used and the down and up sampling processes were embedded within the iterative processes.
Furthermore, to increase the optimization performance, the adequate background value of 37$\%$ in comparison to letter pixel value was added to the object space 
by considering the suppression effect of the high-order noises when using the relatively dense object.
A letter image is fully filled in the vacant space whose size is equal to the hologram magnitude in vertical direction.

A phase hologram prepared at a half of the critical distance, 73.1 mm is displayed in Fig. 7(a).
We found that the pixel values were scattered in the whole area of the digital hologram when compared with those of Fig. 6(b).
The image without interfering the high-order noises is well reconstructed in Fig. 7(b), where the horizontal and vertical pixel resolutions of images are 2.25 µm and 4 µm, respectively.
Figure 7(c) is the unfocused image captured at a distance slightly apart from half of $z_c$.
We observed that an accommodation effect is very strong, which means that the spreading angle of diffractive wave from the image spot would be larger.
Figure 7(d) is the reconstructed image from the phase hologram synthesized at a quarter of the critical distance, 36.5 mm.
Likewise, the hologram reconstructs the image well, without the high-order noises.

Figure 8 is the numerical simulation used to measure the spreading angle of propagating wave from the image.
As shown in Fig. 8(a), when there is no additional process, the diffraction scope of the propagating wave from the circular object image complies with the relation in Eq. (1).
However, the upsampled digital hologram by a simple duplication pixel deteriorates the diffracting behavior from the focused image, as shown in Fig. 8(b).
This is the root cause for the obstruction of the viewing-angle expansion of a holographic image despite the effective removal of the high-order noises.

\section{Viewing-angle enlargement of holographic image and discussions}
As previously stated in Section 2, the replica fringes are formed in the enhanced-NA Fresnel hologram because of the undersampling of the point spread function,
where the finite-sized object could suppress this phenomenon from the near-field diffraction property.
However, it rather produces the adverse effect on the fringe concentration.
To enhance the diffraction performance of a digital hologram by an incident wave, the energy compaction should be mitigated.
A simple method is to diffuse an aggregated hologram pattern by adding the random phase to the real-valued object in the hologram synthesis,
where the restoration of replica fringes is inevitable.

We found that as displayed in Fig. 7(a), the iterative algorithm including the down and up sampling processes effectively prevents the replica fringe formation.
The hologram pattern about the reversely propagating wave seems to be created to compensate the corresponding high-order noises during the repetition processes.
This is different from a spatial filtering appearing in the algorithms with the double Fourier transforms.
However, as seen in Fig. 8, the down and up sampling processes still weakens the viewing-angle expansion of a holographic image.

The Rayleigh-Sommerfeld formula using a definite integral can arbitrarily control the pixel specifications in both planes.
Therefore, there is not a requirement for an unnecessary upsampling process within the iterative algorithm.
As described in Subsection 2.2, the viewing angle dependent on the hologram numerical aperture would be conserved.
Figure 9 is the simulation results using the Riemann integral during the repetition processes.
The phase hologram composed of 512×512 pixels was calculated using the letter object with a 4-µm pixel pitch located at a half of $z_c$.
The fringe compaction is not completely eliminated, and there seems to remain some replica fringes, in Fig. 9(a).
The reconstructed image overlapped with the high-order terms is shown in Fig. 9(b).
The inset figure shows the magnified light intensity propagating from the reconstructed image. 
The diffracting performance of focused image undergoes no deterioration. 
We know that the viewing angle is close to the first-order diffraction angle of 6.7$^\circ$ because the overlapped image is viewable.

Especially, we note that the plane wave incident on the replica fringes forms the high-order images in the direction cosine surface other than the primary image plane.
These images correspond to those optically generated by a pixelated structure of the SLM.
Consequently, the existence of replica fringes might not be meaningful for the high-order image formation because they are created irrespective of replica fringes.
In this circumstance,
the high-order images could be effectively removed through an angular low-pass filter.
As illustrated in Fig. 5(a), the primary image is formed by the carrier wave propagating within the angular bandwidth of $\theta_{\pm1}$.
If an angular stop filter blocking the high-order carrier waves is placed in front of the SLM, the high-order images could not be transmitted \cite{24}.
Here, the whole area of the digital hologram contributes to the primary image formation, which still fulfills the viewing angle variation dependent on the hologram numerical aperture.
For recovering an insufficiency of noise removal, the development of a proper algorithm is required.
Presently, it has been known that the optimization algorithms using a captured image by a camera or deep learning technique demonstrate a considerable improvement in an image fidelity \cite{25,26}.
They would become the proper methods for accomplishing a high-quality image in this type of holographic system.

\section{Conclusions}

We studied the expansion of image space in the enhanced-NA Fresnel holographic display system.
The reconstruction behavior largely depends on angular view range subjected to the synthesis algorithm of the phase Fresnel hologram.
Especially when the viewing angle is restricted to a diffraction angle, a complete image without suffering from the high-order noises is well observed in numerical and optical experiments.
For implementing the holographic display, it is important to develop the method for extending the image space keeping the viewing angle of a holographic image.
Since the viewing angle depends on the hologram numerical aperture in principle, the high-order images could be removed through an angular low-pass filter.
This strategy makes it possible to recover the limitation of the narrow viewing-angle problem due to the smaller pixel pitch of a pixelated modulator,
which would realize the holographic display even using a commercial spatial light modulator.

This work was partially supported by Institute of Information \& Communications Technology Planning \& Evaluation (IITP) grants funded 
by the Korea government (MSIP) (2017-0-00049 and 2021-0-00745)

\,

\begin{figure}
\includegraphics[scale=0.9, trim= 0cm 7cm 0cm 3cm]{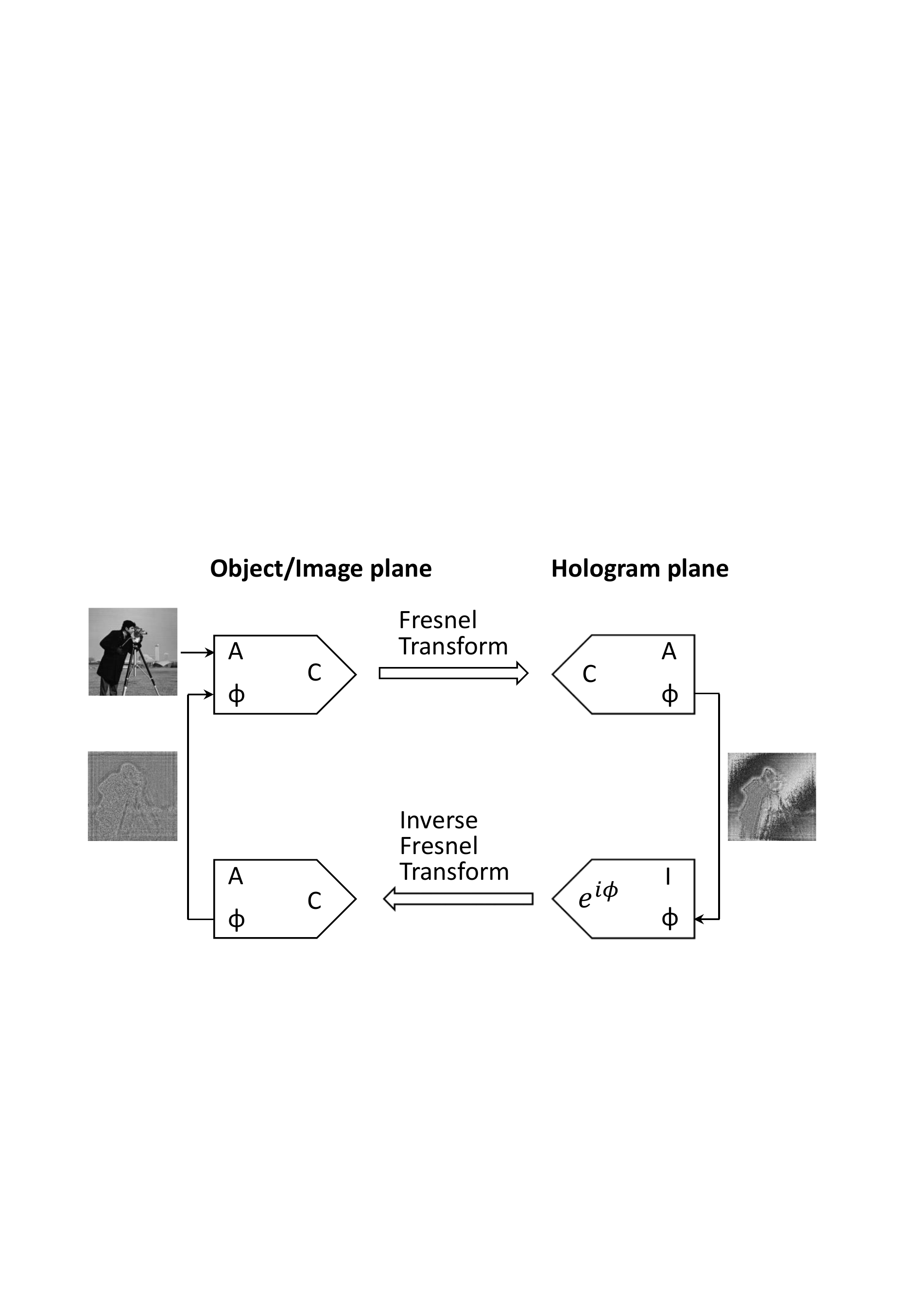}
\caption{Flow chart of the modified Gerchberg–Saxton algorithm to calculate the phase-only hologram. The letters C, A and I indicate complex field, amplitude component,
and the identity value, respectively.}
\end{figure}

\begin{figure}
\includegraphics[scale=0.9, trim= 1.5cm 6.5cm 0cm 2.5cm]{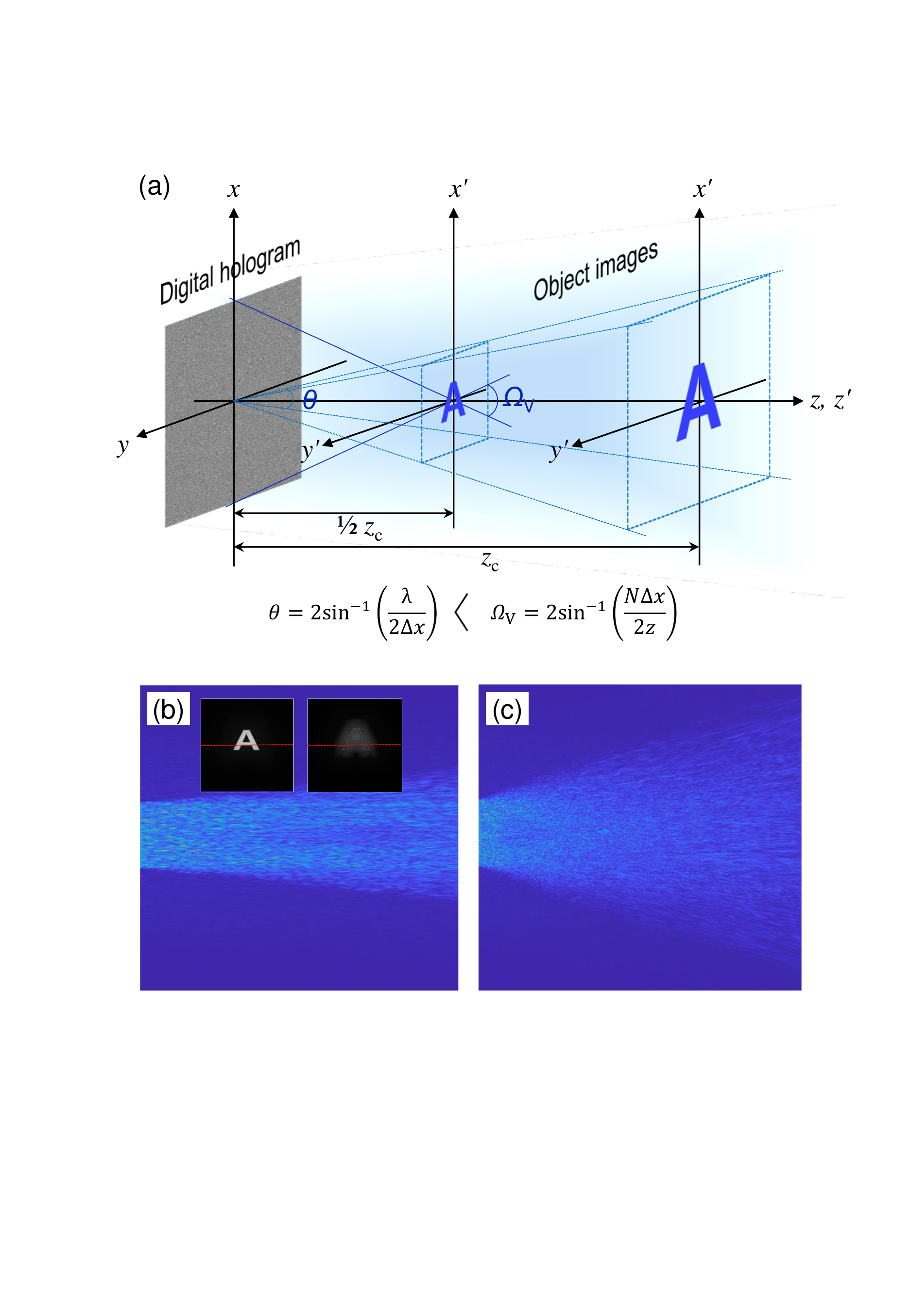}
\caption{Change in the viewing angle of a holographic image reconstructed from the enhanced-NA Fresnel hologram. (a) Schematic diagram of the enhanced-NA Fresnel hologram imaging system.
Intensity profiles of diffractive wave from the focused image reconstructed from the phase holograms synthesized at (b) a critical distance, $z_c$ and (c) a half of $z_c$.
Lateral line images in inset picture are displayed with distance.
A spreading angle of a propagating wave is larger than the diffraction angle by a hologram pixel, when the focused image is placed at a distance lower than a critical distance.}
\end{figure}

\begin{figure}
\includegraphics[scale=0.9, trim= 1.5cm 10cm 0.0cm 1cm]{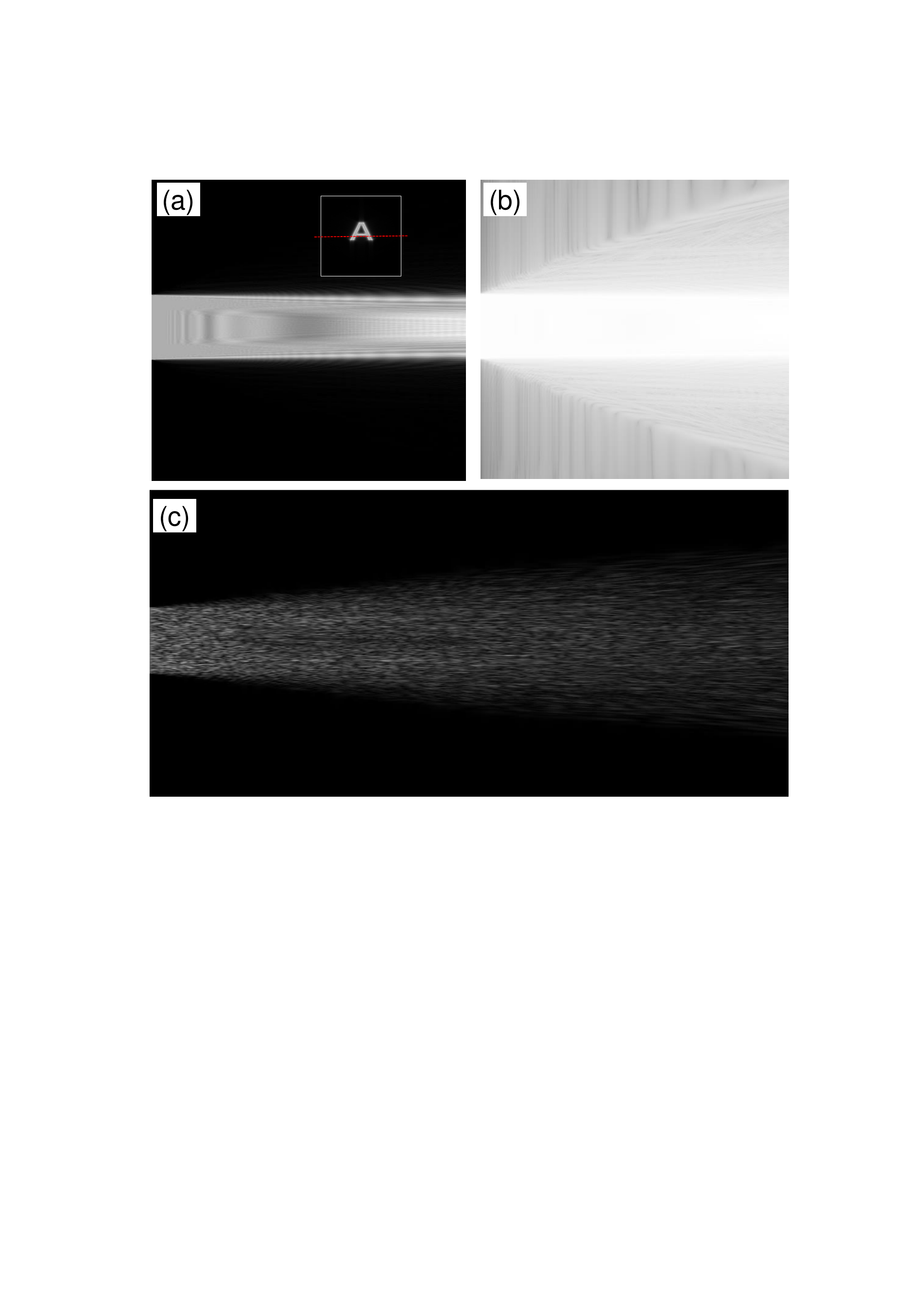}
\caption{Propagating waves of the focused image reconstructed from the directly calculated complex holograms.
Intensity profiles of the diffractive wave in (a) a linear scale and (b) a logarithmic scale, which are simulated using the hologram synthesized at a half of $z_c$.
(c) Diffractive behavior of the image reconstructed from the hologram with no energy compaction.}
\end{figure}

\begin{figure}
\includegraphics[scale=0.85, trim= 0.5cm 8cm 0cm 0cm]{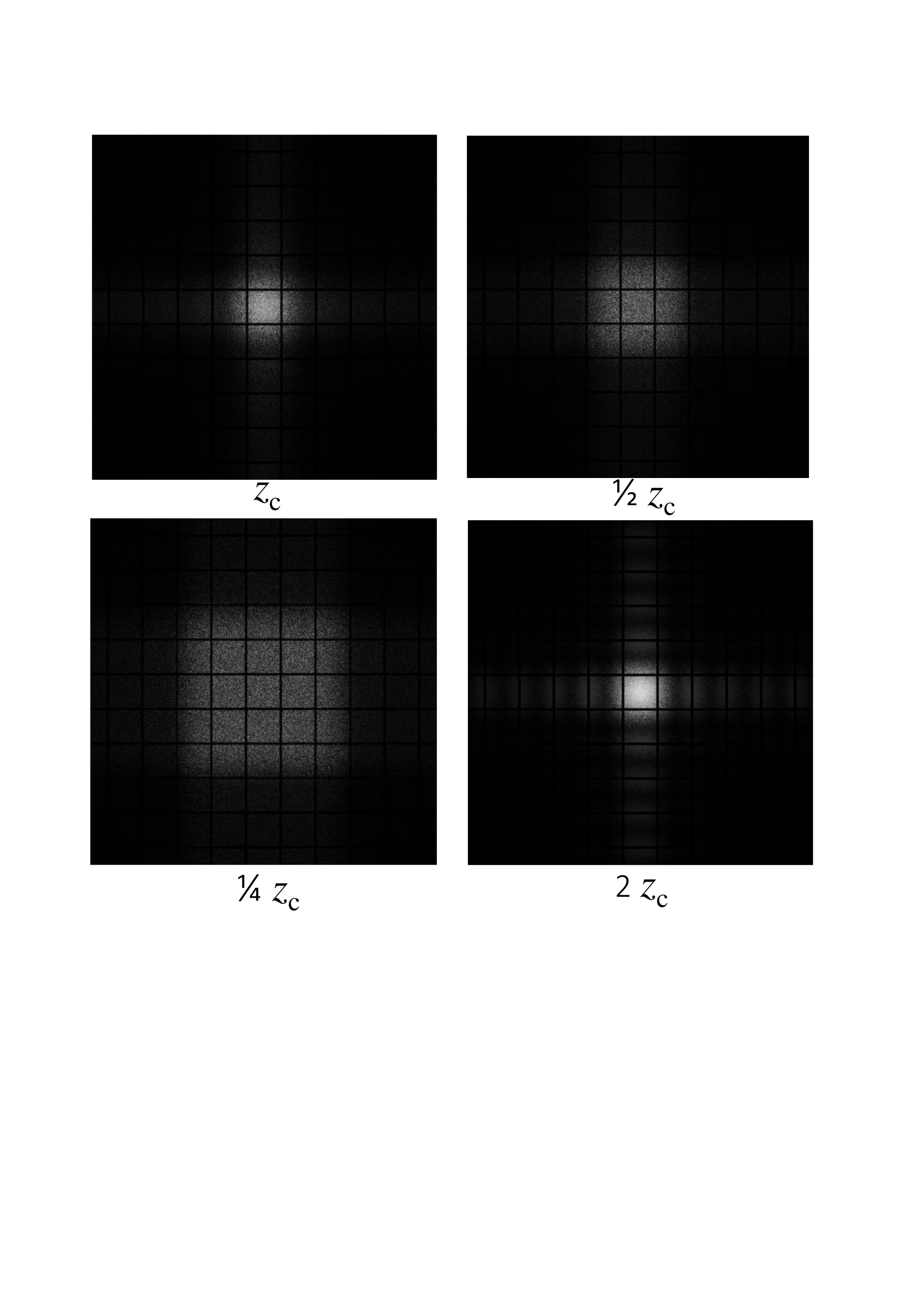}
\caption{Simulation results for the generation of high-order images in the enhanced-NA Fresnel diffraction regime.
A rectangular object consisting of 256×256 pixels has a boundary filled with zeros to distinguish high-order terms.
The bright areas of all the images are of the same size as a digital hologram.}
\end{figure}

\begin{figure}
\includegraphics[scale=1, trim= 1cm 7cm 1cm 2cm]{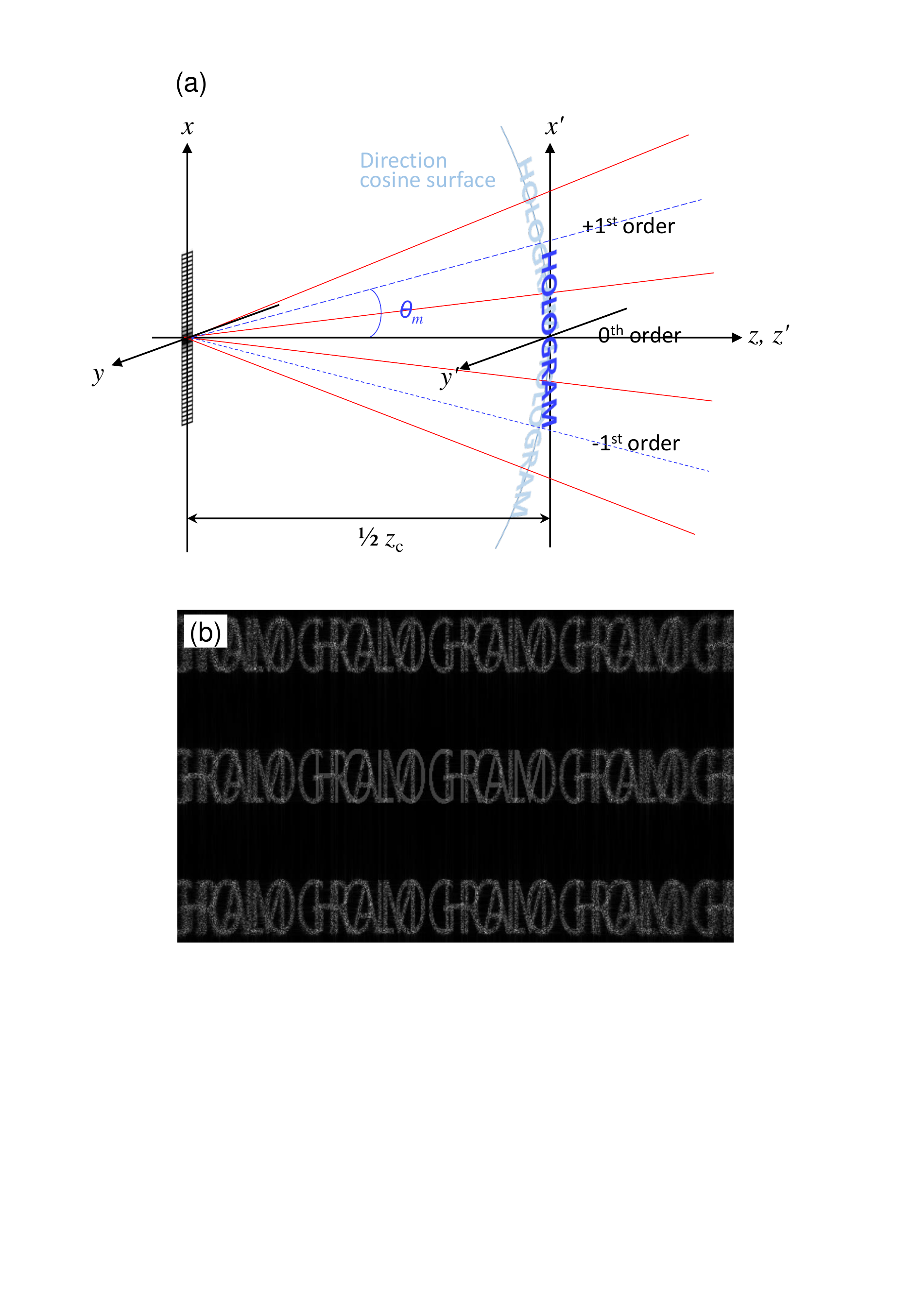}
\caption{Properties of image reconstruction in the extended image space from the enhanced-NA Fresnel hologram. 
A digital hologram is made using the letter image placed at a distance of half of $z_c$.
(a) High-order images are generated along the direction cosine surface at an angular interval, $\theta_{m}$. 
(b) Numerically reconstructed images showing up to the second order are cropped for convenience.}
\end{figure}

\begin{figure}
\includegraphics[scale=0.9, trim= 1cm 6cm 1cm 2cm]{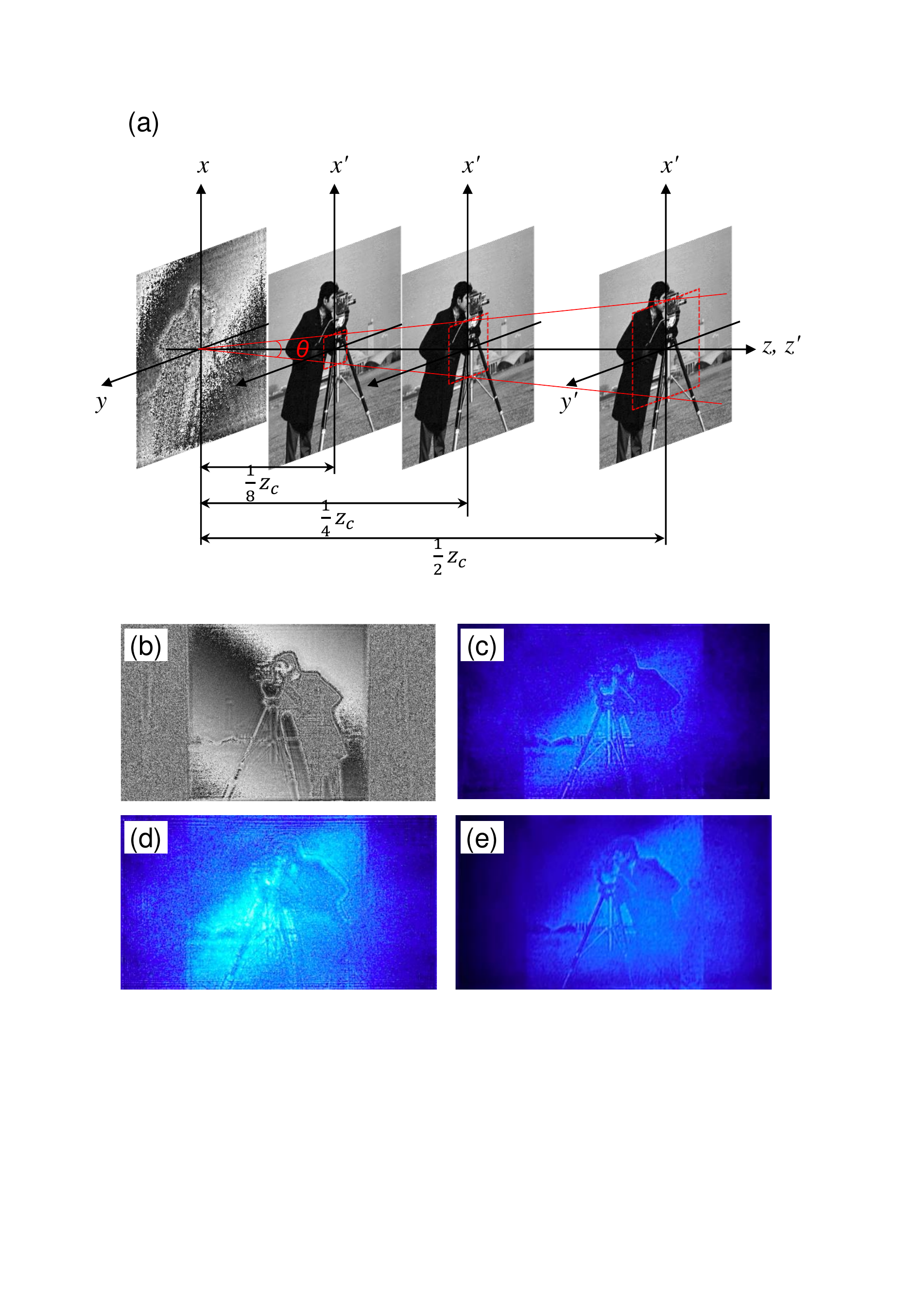}
\caption{Optical reconstruction from the digital hologram made using the extended object space. (a) Phase holograms were prepared at distances of a half, a quarter, and one eighth of $z_c$.
(b) Synthesized digital hologram at half of $z_c$, (c) reconstructed image, and (d) unfocused image at 140 mm. (e) Reconstructed image at quarter $z_c$.}
\end{figure}

\begin{figure}
\includegraphics[scale=1.1, trim= 3cm 14cm 1cm 2cm]{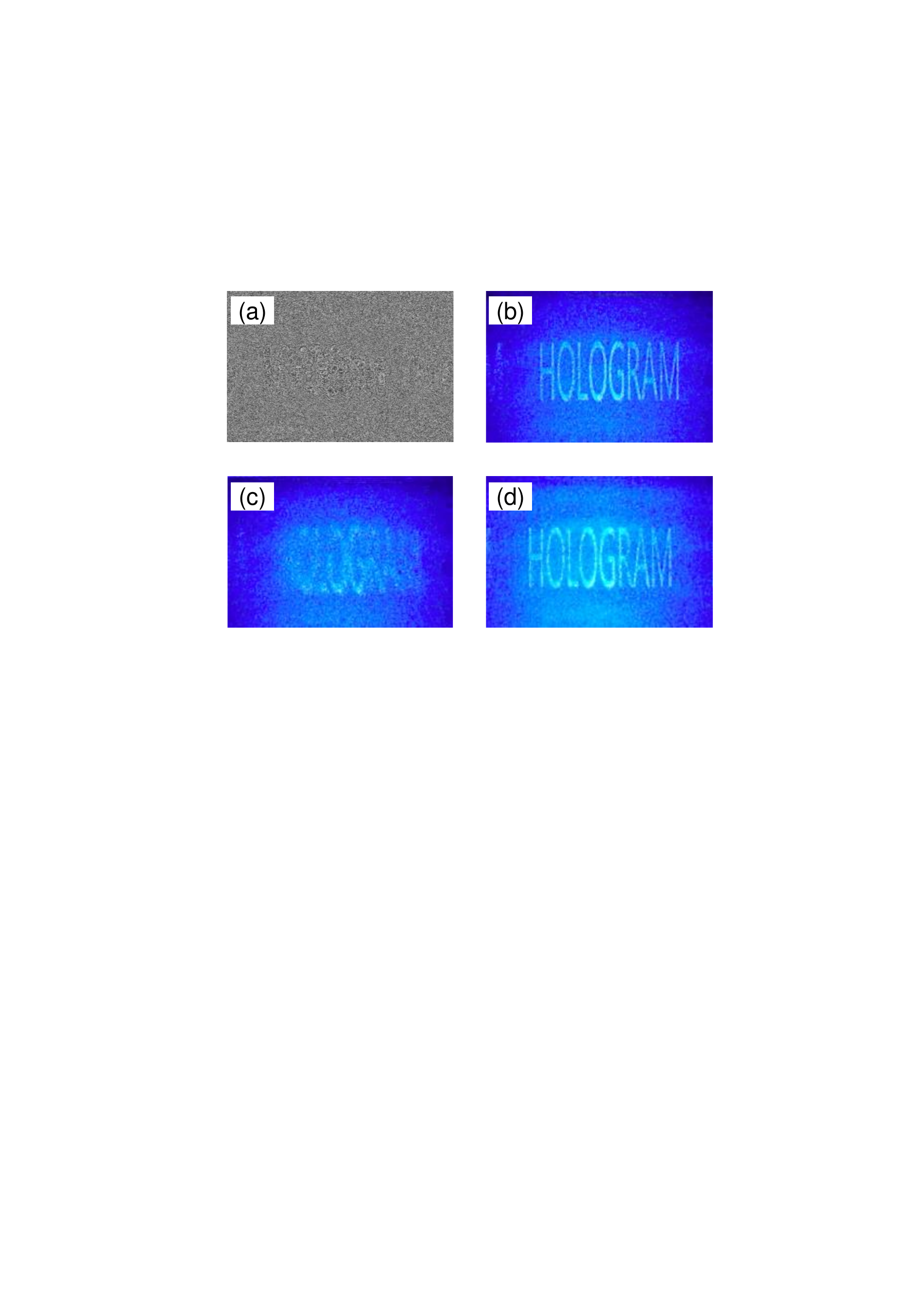}
\caption{Reconstruction property from the enhanced-NA phase hologram synthesized using the sparse letter object.
(a) Phase hologram prepared at a half of critical distance, 73.1 mm, (b) reconstructed image, and (c) unfocused image.
(d) Reconstructed image from the phase hologram made at quarter $z_c$.}
\end{figure}

\begin{figure}
\includegraphics[scale=0.8, trim= 1cm 14cm 1cm 2cm]{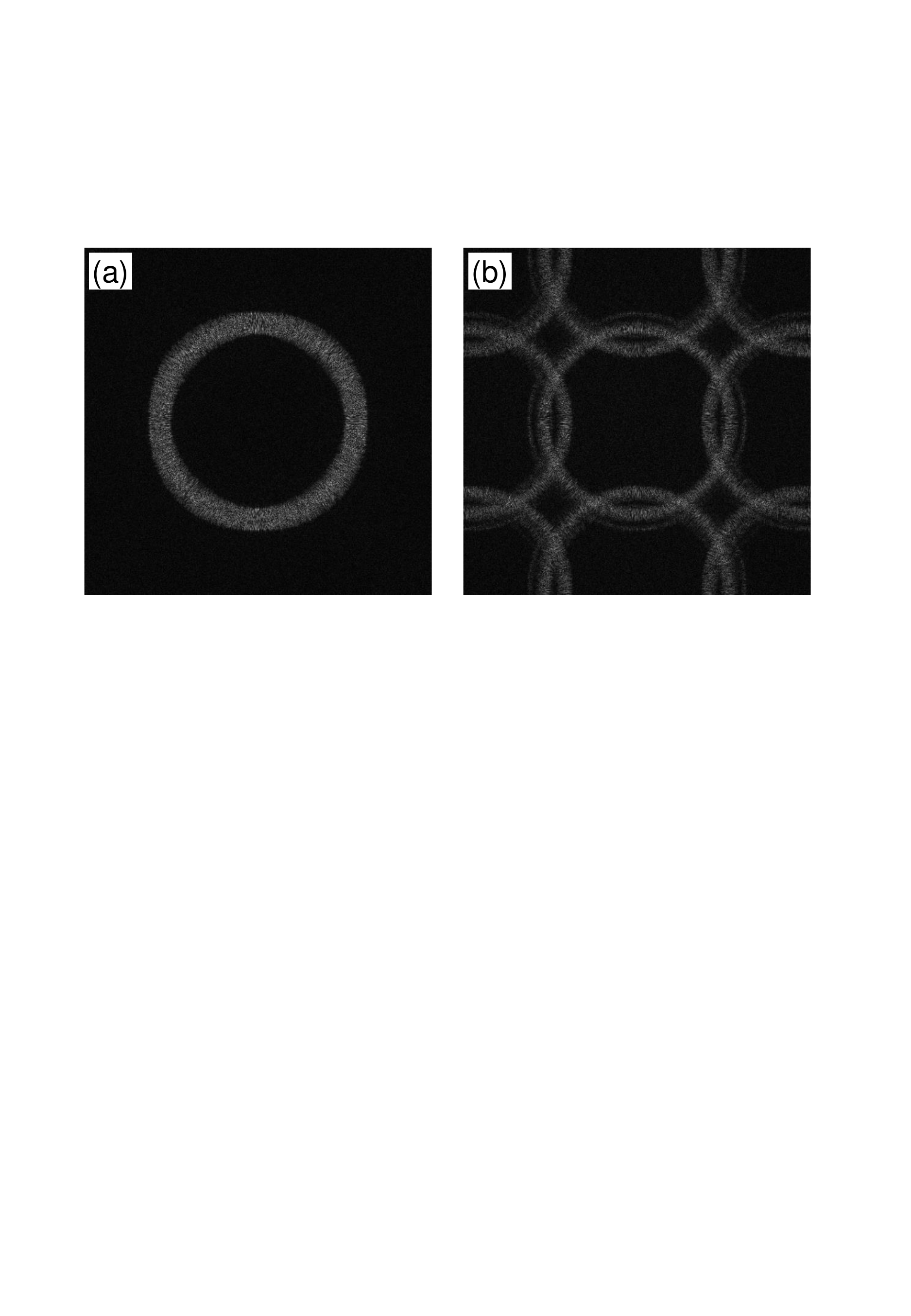}
\caption{Diffraction behavior from the focused image. Digital hologram is prepared using a circular object adding a random phase.
All images are numerically reconstructed at a distance, 78.1 mm.
(a) The diffractive wave from the circular object image spreads out well, whereas (b) the upsampled digital hologram by a simple duplication pixel deteriorates the diffracting behavior.}
\end{figure}

\begin{figure}
\includegraphics[scale=0.9, trim= 1cm 15cm 1cm 2cm]{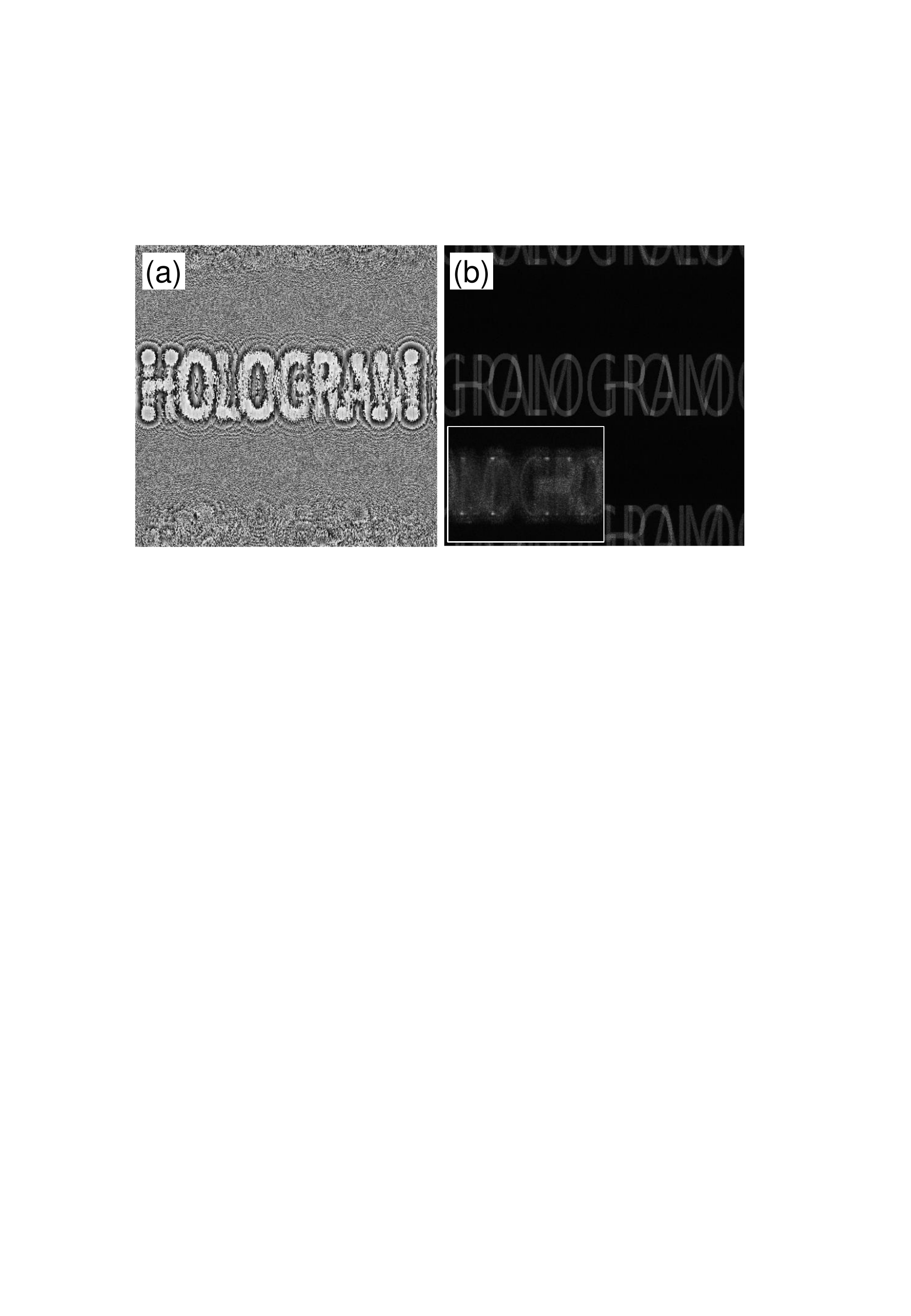}
\caption{Reconstruction results using the Riemann integral during the repetition processes. (a) Synthesized hologram and (b) reconstructed image.
The inset shows the magnified light intensity propagating from the reconstructed image.}
\end{figure}


\begin{thebibliography}{99}

\bibitem{1} J. W. Goodman, Introduction to Fourier Optics (McGraw-Hill, 1996).
\bibitem{2} J. Park, K. Lee, and Y. Park, “Ultrathin wide-angle large-area digital 3D holographic display using a non-periodic photon sieve," Nat. Commun. $\bf{10}$, 1304 (2019).
\bibitem{3} K. Yamamoto, Y. Ichihashi, T. Senoh, R. Oi, and T. Kurita, “3D objects enlargement technique using an optical system and multiple SLMs for electronic holography,” Opt. Express $\bf{20}$(19), 21137-21144 (2012).

\bibitem{4} P. W. M. Tsang and T.-C. Poon, “Novel method for converting digital Fresnel hologram to phase-only hologram based on bidirectional error diffusion,” Opt. Express $\bf{21}$(20), 23680-23686 (2013).
\bibitem{5} J. Hahn, H. Kim, Y. Lim, G. Park, and B. Lee, “Wide viewing angle dynamic holographic stereogram with a curved array of spatial light modulators,” Opt. Express $\bf{16}$(16), 12372-12386 (2008).
\bibitem{6} T. Kozacki, M. Kujawińska, G. Finke, B. Hennelly, and N. Pandey, “Extended viewing angle holographic display system with tilted SLMs in a circular configuration,” Appl. Opt. $\bf{51}$(11), 1771-1780 (2012).
\bibitem{7} Y. Takaki and S. Uchida, “Table screen 360-degree three-dimensional display using a small array of high-speed projectors,” Opt. Express $\bf{20}$(8), 8848-8861 (2012).
\bibitem{8} F. Yaras, H. Kang, and L. Onural, “Circular holographic video display system,” Opt. Express $\bf{19}$(10), 9147-9156 (2011).

\bibitem{9} B. G. Chae, “Wide viewing-angle holographic display based on enhanced-NA Fresnel hologram,” Opt. Express $\bf{29}$(23), 38221-38236 (2021).
\bibitem{10} B. G. Chae, “Analysis on angular field of view of holographic image dependent on hologram numerical aperture in holographic display,” Opt. Eng. $\bf{59}$(3), 035103 (2020).

\bibitem{11}	L. Onural, “Sampling of the diffraction field," Appl. Opt. $\bf{39}$(32), 5929-5935 (2000).
\bibitem{12}	A. Stern and B. Javidi, “Analysis of practical sampling and reconstruction from Fresnel fields," Opt. Eng. $\bf{43}$(1), 239-250 (2004).
\bibitem{13} B. G. Chae, “Analysis on image recovery for on-axis digital Fresnel hologram with aliased fringe generated from self-similarity of point spread function,” Opt. Commun. $\bf{466}$, 125609 (2020).

\bibitem{14} F. Wyrowski and O. Bryngdahl, “Iterative Fourier-transform algorithm applied to computer holography,” J. Opt. Soc. Am. A $\bf{5}$(7), 1058-1065 (1988).
\bibitem{15} C. K. Hsueh and A. A. Sawchuk, “Computer-generated double-phase holograms,” Appl. Opt. $\bf{17}$(24), 3874-3883 (1978).
\bibitem{16} Y. Wu, J. Wang, C. Chen, C. Liu, F. Jin, and N. Chen, “Adaptive weighted Gerchberg-Saxton algorithm for generation of phase-only hologram with artifacts suppression,” Opt. Express $\bf{29}$(2), 1412-1427 (2021).
\bibitem{17} R. W. Gerchberg and W. O. Saxton, “A practical algorithm for the determination of phase from image and diffraction plane pictures,” Optik $\bf{35}$, 237-246 (1972).
\bibitem{18} J. R. Fienup, “Phase retrieval algorithms: a comparison,” Appl. Opt. $\bf{21}$(15), 2758-2769 (1982).

\bibitem{19} D. Mas, J. Garcia, C. Ferreira, L. M. Bernardo, and F. Marinho, “Fast algorithms for free-space diffraction patterns calculations,” Opt. Commun. $\bf{164}$, 233-245 (1999).

\bibitem{20} D. M. Cottrell, J. A. Davis, T. Hedman, and R. A. Lilly, “Multiple imaging phase-encoded optical elements written as programmable spatial light modulators,” Appl. Opt. $\bf{29}$(17), 2505-2509 (1990).
\bibitem{21} E. Carcole, J. Campos, I. Juvells, and S. Bosch, “Diffraction efficiency of low-resolution Fresnel encoded lenses,” Appl. Opt. $\bf{33}$(29), 6741-6746 (1994).
 
\bibitem{22} M. G. Moharam and T. K. Gaylord, “Three-dimensional vector coupled-wave analysis of planar-grating diffraction,” J. Opt. Soc. Am. $\bf{73}$(9), 1105–1112 (1983).
\bibitem{23} J. E. Harvey and C. L. Vernold, “Description of diffraction grating behavior in direction cosine space,” Appl. Opt.  $\bf{37}$, 8158–8160 (1998).

\bibitem{24} B. G. Chae, “Digital holographic display apparatus,” KR patent application 10-2022-0069330 (2022).

\bibitem{25} Y. Peng, S. Choi, N. Padmanaban, and G. Wetzstein, “Neural holography with camera-in-the-loop training,” ACM Trans. Graph. $\bf{39}$(6), 185:1-185:14 (2020).
\bibitem{26} L. Shi, B. Li, C. Kim, P. Kellnhofer, and W. Matusik, “Towards real-time photorealistic 3D holography with deep neural networks,” Nature $\bf{591}$, 234-239 (2021).


\end{thebibliography}
\end{document}